\documentclass[12pt]{iopart}
\usepackage{iopams}
\usepackage[breaklinks=true,colorlinks=true,linkcolor=blue,urlcolor=blue,citecolor=blue]{hyperref}
\usepackage[english]{babel}
\usepackage{setspace}  
\usepackage{graphicx}
\usepackage{amssymb}
\usepackage{amscd}
\usepackage{rotating}
\usepackage{color}
\expandafter\let\csname equation*\endcsname\relax
\expandafter\let\csname endequation*\endcsname\relax
\usepackage{amsmath}

\newcommand{\barr}{\begin{eqnarray}}
\newcommand{\earr}{\end{eqnarray}}
\newcommand{\stdif}[2]{\frac{\partial #1}{\partial #2}}

\begin{document}

\title{Approximate expression for the ground-state energy of the two- and three-dimensional Hubbard model at arbitrary filling obtained from dimensional scaling}

\author{L. N. P. Vilela$^1$, K. Capelle$^{2,3}$, L. N. Oliveira$^4$, V. L. Campo$^5$}

\address{$\mbox{}^1$Laborat\'orio Nacional de Luz S\'incrotron, Campinas, SP, Brazil}
\address{$\mbox{}^2$Laborat\'orio Nacional de Pesquisa em Materia e Energia (CNPEM), Campinas, SP, Brazil}

\address{$\mbox{}^3$Centro de Ci\^encias Naturais e Humanas, Universidade Federal do ABC (UFABC), Rua Santa Ad\'elia, 166, Bairro Bangu, 09210-107, Santo Andr\'e, SP, Brazil}

\address{$\mbox{}^4$Departamento de F\'isica e Inform\'atica, Instituto de F\'isica de S\~ao Carlos, Universidade de S\~ao Paulo, Caixa Postal 369, 13560-970, S\~ao Carlos, SP, Brazil}
\ead{vivaldo.leiria@gmail.com, vlcampo@df.ufscar.br}

\address{$\mbox{}^5$Departamento de F\'isica, Universidade Federal de S\~ao Carlos, Rodovia Washington Lu\'is, km 235, Caixa Postal 676, 13565-905, S\~ao Carlos, SP, Brazil}

\date{\today}

\begin{abstract}
We generalize the linear discrete dimensional scaling approach for the repulsive Hubbard model to obtain a nonlinear scaling relation that yields accurate approximations to the ground-state energy in both two and three dimensions, as judged by comparison  to auxiliary-field quantum Monte Carlo (QMC) data. Predictions are made for the per-site ground-state energies in two and three dimensions for $n$ (filling factor) and $U$ (Coulomb interaction) values for which QMC data are currently unavailable.
\end{abstract}
\maketitle

\section{Introduction}  
\label{intr}

Scaling relations pervade physics. In their simplest form, they arise, e.g., in dimensional analysis, as formalized by Buckingham's theorem \cite{pitheo}, according to which quantities pertaining to one of two geometrically and dynamically similar systems can be inferred by measuring or computing the corresponding quantity in the other, if the basic variables are suitably scaled.

Similar, but more sophisticated, scaling relations appear in statistical physics, for example in the context of universality. In critical phenomena \cite{cardybook}, the dependence of some property $P$ on the temperature near the critical point, expressed in multiples of an appropriate unit $P_u$, is given by universal scaling functions of $T/T_C$, where $T_C$ is the critical temperature,
\begin{align}\label{screl_tc}
\frac{P}{P_u} = f_{univ}\left(\frac{T}{T_C}\right). 
\end{align}
Different physical systems in the same universality class can have different critical temperatures, but for the same ratio $T/T_C$, they present the same universal properties.

In Ref.~\cite{LK2006}, a  different kind of scaling, denoted {\it discrete dimensional scaling} (DDS), was identified. DDS relates dimensionless ratios computed in systems differing in their dimensionality $d$. 

Systems of different dimensionalities are normally considered to be completely distinct objects. In statistical mechanics, systems with different dimensionalities belong to different universality classes, characterized by different critical exponents \cite{StanleyRMP,FisherRMP}. In condensed-matter physics, one-, two- and three-dimensional (1D,2D,3D) systems are known to display very different phenomenology. This behaviour is reflected by commonly used model Hamiltonians, such as the repulsive Hubbard model (HM) which in 1D displays a Luttinger-liquid and a Mott-insulator phase, in 3D follows the Fermi-liquid paradigm, and in 2D has a complex ground-state that is not fully understood, but characterized by strong antiferromagnetic correlations and possible superconducting and charge fluctuations \cite{tasaki98,Essler}. Little, if any, important physics remains the same on crossing the dimensional divide.

On the other hand, remarkable progress has been made in specific situations by exploiting relations between systems and quantities in different dimensionalities. In statistical physics, for example, the $\epsilon$-expansion \cite{WilsonEpsilon}  treats $d$ as a continuous variable and expands around $d=4$. In condensed-matter physics, dynamical-mean-field theory (DMFT) treats many-body models in physical dimensionalities by starting from the limit $d\to\infty$ \cite{dmft}.

The enormous success of the $\epsilon$-expansion and of DMFT in their respective domains notwithstanding, there is little work on systematically exploring connections between different dimensionalities.  However, in addition to these two concrete examples, general mathematical considerations suggest that such interdimensional relations may be more ubiquitous than usually thought. 

Nonrelativistic quantum systems are described by Schrö\-din\-ger's equation, a linear partial differential equation (PDE). The behaviour of PDEs and their solutions is determined by their dimensionality $d$, in conjunction with several other defining features, such as its characteristics (elliptic, parabolic, hyperbolic or mixed), order (highest derivative), degree (power of the highest derivative), linearity or nonlinearity (power of the solution function), and whether it has constant or variable coefficients, and a real or complex solution space, among others \cite{TikhonovSamarski}. The Schrödinger equations defined by a Hamiltonian in dimensionalities $d$ and $d'$ are the same in all these key aspects, except for $d$ itself. This strongly suggests that a system in $d$-dimensions cannot be arbitrarily different from its counterpart in $d'$-dimensions and that some connection between their properties must exist.

A more intuitive way of thinking about this connection is to note that low-dimensional systems are the basic building blocks out of which higher-dimensional systems are composed: sites connect to form chains, parallel chains define a plane and stacked planes a three-dimensional lattice. Again, some features of the building blocks should be expected to be preserved in the constituting system.

The quantum-mechanical virial theorem for bound Coulomb systems~\cite{grossbook} provides a beautiful example: since in any dimension $d$ the total energy is equal to the negative of the kinetic energy, $E(d) = -T(d)$, we have the dimensional scaling relation
\begin{align}
\frac{E(d)}{T(d)} = \frac{E(d')}{T(d')},\label{screl2}
\end{align}
which is a first example of DDS, as it relates a ratio in dimension $d$ to the same ratio in dimension $d'$. 

For the Fermi liquid, relation (\ref{screl2}) holds exactly, for any $d$, $d'$ and density $n$, due to the virial theorem. Interestingly, the same relation also holds in two limiting cases of the Hubbard model, as it is by construction exact at $U=0$ and, due to the Hellmann-Feynman theorem, also at $U\to\infty$. Thus,  Eq.~(\ref{screl2}) provides a concrete example of how physical or mathematical constraints (the virial theorem and the Hellmann-Feynman theorem, respectively) establish connections between quantities in different dimensionalities.

The basic linear relation (\ref{screl2}) can be rewritten as
\begin{align}
E(d) = \frac{E(d')}{T(d')} T(d) = \frac{T(d)}{T(d')} E(d') \approx \frac{T_s(d)}{T_s(d')} E(d') ,
\label{screl2b}
\end{align}
where $T$ denotes interacting and $T_s$ noninteracting kinetic energies, which are easier to calculate. In this latter form, dimensional scaling becomes a tool for approximating the ground-state energy in $d$-dimensions  in terms of the ground-state energy in $d'$-dimensions by scaling the latter by a simple energy ratio.

What makes this observation potentially useful in studying ground-state properties, is that numerical calculations for the Hubbard model have shown that Eq.~(\ref{screl2b}), even in its version employing noninteracting kinetic energies, provides fair approximations also for intermediate values of $U$ (including the physically realistic range), both at $n=1$ \cite{LK2006} and for $n \neq 1$ \cite{akandeBJP}.

This suggests that we use dimensional scaling to build approximations for physical quantities in a dimensionality in which calculations are easier to perform and scale them to a different, harder-to-describe, dimensionality. Of course, the difference between the physics in the two involved dimensionalities must be approximately accounted for by the scaling factor, which cannot always be expected to be of the simple form in Eq.~(\ref{screl2b}). 

Motivated by these considerations, we develop, in the present paper, nonlinear generalizations of (\ref{screl2}) and (\ref{screl2b}) for the Hubbard model, which are of a general form inspired by Eq.~(\ref{screl_tc}):
\begin{align}
\frac{E(d)}{T_s(d)} = f_{d,d'}\left(\frac{E(d')}{T_s(d')}\right).\label{screl3}
\end{align}

We start by providing, in Sec.~\ref{seclds}, a justification for the previously \cite{LK2006} {\it ad hoc} replacement of the interacting by the non-interacting kinetic energy on the right-hand side of Eq.~(\ref{screl2b}). In section~\ref{secnlds}, we then construct the corresponding nonlinear dimensional scaling relation for the half-filled model and for arbitrary filling. 
Our main result is an explicit approximation for the scaling function $f_{d,d'}\left(E(d')/T_s(d')\right)$, with $d'=1$, which we use to calculate ground-state energies for $d=2$ and $d=3$.
A detailed comparison with quantum Monte Carlo (QMC) data in the literature shows that the nonlinear generalization significantly improves the accuracy of scaling with noninteracting kinetic energies, both at $n=1$ and for $n\neq 1$, for physically realistic values of $U$. The nonlinear scaling relation is simple enough to allow us to make predictions for the per-site ground-state energy of the two- and three-dimensional Hubbard model in parameter ranges that have not yet been explored by QMC, and where Monte Carlo techniques may be hard to apply.

\section{Linear dimensional scaling}
\label{seclds}

The Hubbard model Hamiltonian reads
\begin{align}
H = -t\sum_{\langle i,j\rangle, \sigma}~\left (c^\dagger_{i,\sigma} c_{j,\sigma} + c^\dagger_{j,\sigma} c_{i,\sigma} \right) + U\sum_i \hat{n}_{i,\uparrow}\hat{n}_{i,\downarrow},\label{hubmod}
\end{align}
where $c_{i,\sigma}^\dagger$ creates a fermion on site $i$ with spin $\sigma$ $(\uparrow,\downarrow)$. The first sum in (\ref{hubmod}) runs over all pairs of first neighbors and represents the kinetic energy. The hopping amplitude $t$, which sets the kinetic-energy scale, will define our unit of energy. For dimensionalities $d=2$ or 3, we will consider the model on a square or cubic lattice, respectively. In the second sum, $\hat{n}_{i,\sigma} = c_{i,\sigma}^\dagger c_{i,\sigma}$, so that any doubly occupied site adds the intra-site interaction $U$ to the system energy. We are interested in repulsive interactions ($U>0$). 

In one dimension, this model is exactly solvable by the Bethe Ansatz~\cite{liebwu,schlott}. Away from half-filling, it is an example of a Luttinger liquid~\cite{giamarchibook}, with spin-charge separation. At half-filling, the system is a Mott insulator. An energy gap separates the ground-state from the charge excitations for any $U>0$~\cite{liebwu}. For $d>1$, no exact solution is known. The most reliable information about the properties of the model has been obtained numerically, mostly from QMC calculations \cite{prx15}.

\begin{figure}[t]
\centerline{\includegraphics[width=0.95\columnwidth]{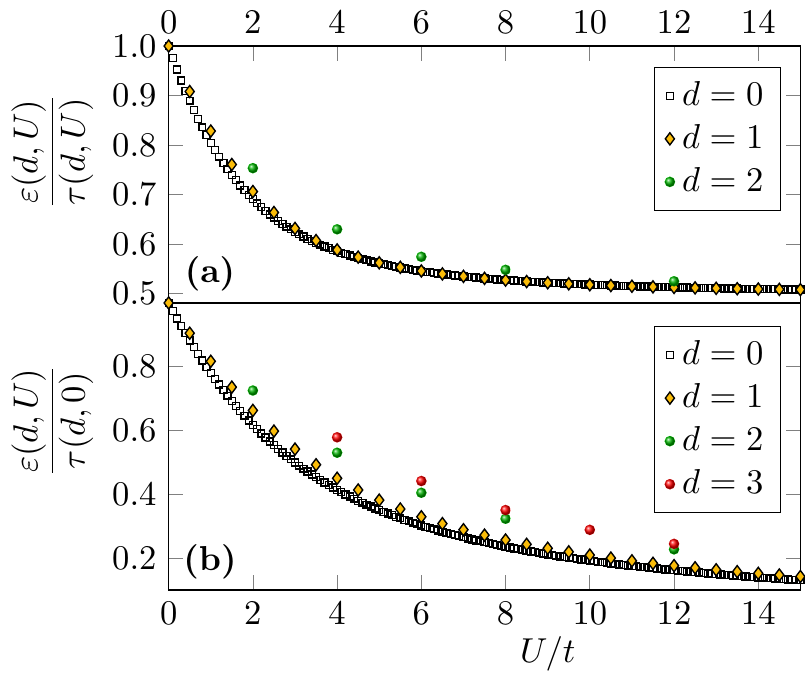}}
\caption{\label{fig1} Ratios (a) $\epsilon(d,U)/\tau(d,U)$ and (b) $\epsilon(d,U)/\tau(d,0)$ between the total energy per site and the interacting and non-interacting kinetic energy per site, respectively, for the half-filled $d$-dimensional Hubbard model with $d=0,1,2,3$. The green balls representing $d=2$ come from the AFQMC numerical data in Ref.~\cite{prx15}. The red balls representing $d=3$ in panel (b) come from Ref.~\cite{vmc-shibaII}; panel (a) shows no corresponding points because accurate results for $\tau(3,U)$ are unavailable.
}
\end{figure}

\subsection{Half-filled case}
\label{sec:half-filled-case-1}
The dimensional scaling approach was originally introduced~\cite{LK2006} as a procedure to obtain approximations to the ground-state energy of the half-filled Hubbard model in 2 and 3 dimensions by evaluating the approximate linear scaling relation
\begin{align}
\frac{\epsilon(d,U)}{\tau(d,U)} \approx \frac{\epsilon^{\rm approx}(d,U)}{\tau(d,U)} \equiv \frac{\epsilon(d',U)}{\tau(d',U)}, \label{linear0} 
\end{align}
for $d'=1$. Here, $\epsilon(d,U)$ is the ground-state energy per site for the $d$-dimensional model with intra-site interaction $U$ at half-filling, $\epsilon^{\rm approx}(d,U)$ is an approximation to it defined by this equation, and 
$\tau(d,U)$ is the interacting kinetic energy per site. 

As a direct consequence of the Hellmann-Feynman theorem, the kinetic energy can be calculated from the full ground-state energy as
\begin{align}
\tau(d,U) = \epsilon(d,U) - U\stdif{\epsilon(d,U)}{U}. 
\label{kin}
\end{align}

The energies in Eqs.~(\ref{linear0}) and (\ref{kin}) can be computed exactly for $d\to0$ (approximated as the smallest possible Hubbard model, namely a Hubbard dimer with open boundary condition) and for $d=1$. In particular, for $d=1$ at half-filling \cite{liebwu}
\begin{align}
\epsilon(1,U) = -4\int_0^\infty\frac{J_0(x)J_1(x)}{x\left[1 + e^{Ux/2}\right]}~dx,
\end{align}
where $J_0$ and $J_1$ are zero and first-order Bessel functions.

Figure~\ref{fig1}(a) displays the ratios $\epsilon(d,U)/\tau(d,U)$ for $d=0,1,2$ as functions of the ratio $U/t$, computed independently. For $d=0,1$, the calculation can be done analytically; for $d=2$ we have resorted to numerical results from auxiliary-field quantum Monte Carlo (AFQMC) calculations reported in Ref.~\cite{prx15}. From the weak dependence on the dimensionality $d$ in Fig.~\ref{fig1}, we conclude that the simple scaling relation (\ref{linear0}) is a fairly satisfactory approximation. Clearly, however, there is room for improvement.

Given $\epsilon(d',U)$ for $d'=0$ or $d'=1$, and $\tau(d, U)$, Eq.~ (\ref{linear0}) yields $\epsilon(d,U)$ for $d>1$.  Unfortunately, this approximation has no practical value, since $\tau(d,U)$ is only known in special limits. Computing the interacting kinetic energy proves as difficult as evaluating the total energy. 

To overcome this limitation, Ref.~\cite{LK2006} proposed a further approximation:  to substitute the noninteracting kinetic energies $\tau(d,0)$ for $\tau(U,d)$ in the scaling relation. A simpler, albeit somewhat less accurate expression resulted:
\begin{align}
\frac{\epsilon(d,U)}{\tau(d,0)} \approx \frac{\epsilon(d',U)}{\tau(d',0)}. \label{linear01} 
\end{align}

Compared to Fig.~\ref{fig1}(a), the plots of the ratios $\epsilon(d,U)/\tau(d,0)$ for $d=0,1,2$ in Fig.~\ref{fig1}(b) show more pronounced dependence on dimensionality. Nonetheless, the green balls representing the ratio for $d=2$, and even the red balls representing $d=3$ are still fairly close to the corresponding black open squares ($d=0$) 0 or orange rhombuses ($d=1$). The separations are smaller than one would expect from the apparently crude approximation converting Eq.~(\ref{linear0}) into Eq.~(\ref{linear01}). Moreover, as we shall see, that approximation warrants good agreement with Quantum Monte Carlo data describing the $U$ dependence of $\epsilon(U,d)$ for $d=2$ and $3$. Brief discussion of the connection between Eqs.~(\ref{linear0})~and  (\ref{linear01}) seems therefore appropriate.

One can easily see that Eq.~(\ref{linear01}), like Eq.~(\ref{linear0}), become exact in the $U=0$ and $U\to\infty$ limits.  $U=0$ makes the numerator on each side of Eq.~(\ref{linear01}) equal to the denominator. For $U\to\infty$, the numerator on each side vanishes~\cite{nagaoka66,takahashi77}, while the denominator is nonzero.

For finite $U$, a more elaborate argument is required. Without sacrificing generality, we may write the interacting energy in $d$-dimensions as a function of the ratio between the one-dimensional energy and the noninteracting one-dimensional energy:
\begin{align}\label{eq:7}
  \epsilon(d,U) = \tau(d, 0)f_d\left(x(U)\right),
\end{align}
with the shorthand
\begin{align}\label{eq:8}
  x(U)\equiv \frac{\epsilon(1,U)}{\tau(1,0)}.
\end{align}

The subindex $d$ on the right-hand side of Eq.~(\ref{eq:7}) is a reminder that the function $f_d$ depends on the dimensionality. Clearly, $f_1(x)\equiv x$.

Differentiate, then, both sides of Eq.~(\ref{eq:7}) with respect to $U$, and divide each side of the resulting equality by the corresponding side of Eq.~(\ref{eq:7}). The following expression results:
\begin{align}\label{eq:3}
  \frac{\partial\epsilon (d,U)}{\partial U} =
 \dfrac{\epsilon(d,U)}{\tau(1,0)f_d\big(x(U)\big)} \dfrac{df_d}{dx}\frac{\partial \epsilon(1,U)}{\partial U}. 
\end{align}

Substitution of the right-hand side for $\partial\epsilon(d,U)/\partial U$ on the right-hand side of Eq.~(\ref{kin}) then yields the equality
\begin{align}\label{eq:9}
  \tau(d,U) = \epsilon(d,U)\left(1- U  \dfrac{1}{\tau(1,0)f_d\big(x(U)\big)} \dfrac{df_d}{dx}\frac{\partial \epsilon(1,U)}{\partial U}\right), 
\end{align}
and multiplication of both sides by $\epsilon(1,U)$ followed by straightforward manipulation shows that
\begin{align}\label{eq:10}
  \dfrac{\epsilon(d,U)}{\tau(d,U)}=
  \dfrac{\epsilon(1,U)}{\epsilon(1,U)- U  \dfrac{\partial \epsilon(1,U)}{\partial U}y_d(U)}
\end{align}
where
\begin{align}\label{eq:11}
  y_d(U) \equiv \dfrac{x(U)}{f_d\big(x(U)\big)} \dfrac{df_d}{dx}.
\end{align}

Equation~(\ref{eq:10}) is formally exact\textemdash formally, since the function $f_d(x(U))$ is unknown, except at $d=1$. To exploit it, we need an approximate form for $f_d(x)$ ($d>1$), which must satisfy two boundary conditions: $f_d(0)=0$, since $\epsilon(d,U\to\infty)=0$, and $f_d(1)=1$, since $\epsilon(d,U=0)=\tau(d,0)$. The simplest approximation consistent with those conditions amounts to assuming $f_d(x)$ independent of $d$:
\begin{align}\label{eq:12}
  f_d\big(x(U)\big) \approx x(U),
\end{align}
which makes Eq.~\eqref{eq:7} equivalent to Eq.~\eqref{linear01}.

Under this approximation, Eq.~\eqref{eq:11} reduces to $y_d(U) = 1$, and the denominator on the right-hand side of Eq.~\eqref{eq:10} reproduces the right-hand side of Eq.~\eqref{kin} for $d=1$. In other words, the approximation $f_d(x)\approx x$ reduces Eq.~\eqref{eq:10} to the $d'=1$ version of Eq.~\eqref{linear0}:
\begin{align}
  \label{eq:1}
  \dfrac{\epsilon(d,U)}{\tau(d,U)} \approx\dfrac{\epsilon(1,U)}{\tau(1,U)}.
\end{align}

Conversely, we can see that Eq.~\eqref{eq:1} implies $y_d(x)=1$. Equation~\eqref{eq:11} then becomes a first-order ordinary differential equation for $f_d(x)$, which can be easily solved under the boundary conditions $f(0)=0$ and $f(1)=1$ to show that $f_d(x)=x$. 

In summary, Eq.~\eqref{kin} implies that Eq.~\eqref{linear01}, with $d'=1$, holds if and only if Eq.~\eqref{eq:1} is valid. The physical arguments in Sec.~\ref{intr} indicate that Eq.~\eqref{linear0} is a good approximation. We therefore expect Eq.~\eqref{linear01} to be a good approximation, which justifies the expedient adopted in Ref.~\cite{LK2006}, which substituted the latter equality for the former.

\begin{figure}[h]
\centerline{\includegraphics[width=0.95\columnwidth]{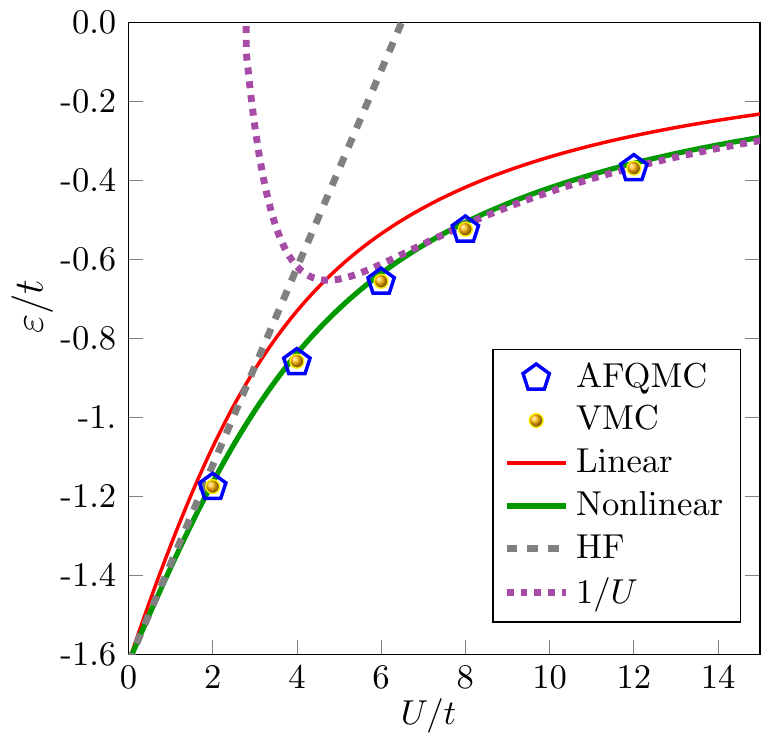}}
\caption{\label{fig2} Per-site ground-state energies for the two-dimensional Hubbard model at half-filling resulting from various approximations. The blue pentagons (auxiliary-field quantum Monte Carlos\textemdash AFQMC) and golden balls (variational quantum Monte Carlo\textemdash VMC) are from Ref.~\cite{prx15}. The violet dashed line ($1/U$ expansion) represents the leading two terms in Takahashi's asymptotic expansion, Eq.~(\ref{largeU}). The dashed gray line represents the Hartree-Fock approximation. The red solid line depicts the linear approximation, Eq.~(\ref{linear01}), and the green solid line depicts Eq.~(\ref{poli}).
}
\end{figure}

One might hope that Eq.~\eqref{linear0} be exact, in which case Eq.~\eqref{linear01} would also be exact. \ref{sec:ineq-betw-line} nonetheless shows that the two equalities are inequivalent: for large $U$, the former is more reliable than the latter. Even without reference to numerical evidence, we can see that the linear approximation~\eqref{eq:12} is incompatible with Eq.~\eqref{kin}\textemdash a corollary of the Helmann-Feynman Theorem. The numerical data, more specifically the separation between the golden balls and the red lines in Figs.~\ref{fig2}~and \ref{fig3}, corroborate this conclusion and offer an estimate of the deviations introduced by the linear approximation~\eqref{eq:12}. This leads us to the discussion in Sec.~\ref{secnlds}, which considers nonlinear approximations. Before that, however, a brief digression is necessary, to recall that the linear approximation has been extended to arbitrary filling \cite{akandeBJP}. 

\subsection{Arbitrary filling}
\label{sec:arbitrary-filling}
To generalize Eq.~\eqref{eq:7}, Ref.~\cite{akandeBJP} let the density (site occupation) $n$ be an additional variable and wrote the expression
\begin{align}
  \epsilon(d,U,n) \approx \epsilon_{\mbox{\tiny L}}(d,U,n) = \tau(d,0,n)x(U,n)
  \qquad(n\le1), \label{linear2n}
\end{align}
where
\begin{align}
x(U,n) = \frac{\epsilon(1,U,n)}{\tau(1,0,n)}. \label{xun}
\end{align}

The restriction $n\le1$ on the right-hand side of Eq.~\eqref{linear2n} is of no practical consequence, for a particle-hole transformation maps the $n$-particle per site problem onto the $2-n$ problem \cite{liebwu}:
\begin{align}
\epsilon(d,U,n) = \epsilon(d,U,2-n) + U(n-1).\label{ph}  
\end{align}

For fixed density $n\leq 1$, in the non-interacting limit, Eq.~\eqref{linear2n} becomes exact, since $x(0,n)=1$. As $U\to\infty$, the per-site ground-state energy for $d=1$ is given by the expression
\begin{align}
\epsilon(1,U\to\infty,n) = \frac{\epsilon(1,0,2n)}{2} = \frac{\tau(1,0,2n)}{2},\label{euinfty} 
\end{align}
which can be physically understood by considering that the limit $U\to \infty$ enforces a special version of the Pauli principle, one that allows single-particle states to be vacant or occupied by at most one electron, with either $s_z=1/2$ or $s_z=-1/2$.
Except for this restriction, the single-particle states are the eigenstates of the $U=0$ Hamiltonian.  In the ground state, each single-particle energy below the Fermi level is singly occupied. The Fermi level for $n$ electrons per site hence coincides with the Fermi level for a noninteracting system with $2n$ electrons per site. The ground-state energy is $\epsilon(1,U\to\infty,n) = -\frac{2}{\pi}\sin(\pi n)$, which implies that
\begin{align}
x(U\to\infty,n) = \cos\left(\frac{\pi n}{2}\right),\label{xuinfty}
\end{align}
so that the interval $[0,\infty)$ for $U$ is mapped onto the interval $(\cos\left(\frac{\pi n}{2}\right),1]$ in the scaling variable $x(U,n)$. 

The simple physical argument following Eq.~(\ref{euinfty}) is valid (i) for any density $n$ at $d=1$~\cite{carmelo88}\footnote{We consider here the thermodynamic limit. For finite chains, only with open boundary condition is Eq.~\eqref{euinfty} correct. To see how periodic boundary condition invalidates the argument, consider a trimer and a particular configuration comprising a spin-up electron at the first site, a vacancy at the second, and a spin-down electron at the third, which can be denoted $|A\rangle=|\uparrow,0,\downarrow\rangle$. We can now easily construct the sequence of states including $|A\rangle$ in which each state is coupled to the following one by the $U\to\infty$ Hamiltonian. Under open boundary condition, the sequence comprises three elements: $|0,\uparrow,\downarrow\rangle$, $|\uparrow,0,\downarrow\rangle$, $|\uparrow,\downarrow,0\rangle$, which match the sequence of spinless-electron states in which each state is coupled to the next by the $U=0$ Hamiltonian: $|0,1,1\rangle$, $|1,0,1\rangle$, $|1,1,0\rangle$. Under periodic boundary condition, the sequence of spinless-electron states is unchanged, but the sequence including state $|A\rangle$ now has six states: $|0,\uparrow,\downarrow\rangle$, $|\uparrow,0,\downarrow\rangle$, $|\uparrow,\downarrow,0\rangle$, $|0,\downarrow,\uparrow\rangle$, $|\downarrow,0,\uparrow\rangle$, $|\downarrow,\uparrow,0\rangle$. The same argument can be extended to lattices of arbitrary sizes and shows that, under periodic condition with $n\ne1$, the spectrum of the $U\to\infty$ Hamiltonian cannot be mapped onto that of the $U=0$ Hamiltonian for spinless electrons.} and (ii) at half-filling for $d>1$~\cite{nagaoka66,shastry90}.

\section{Nonlinear dimensional scaling}  
\label{secnlds}

\begin{figure}[h]
\centerline{\includegraphics[width=0.95\columnwidth]{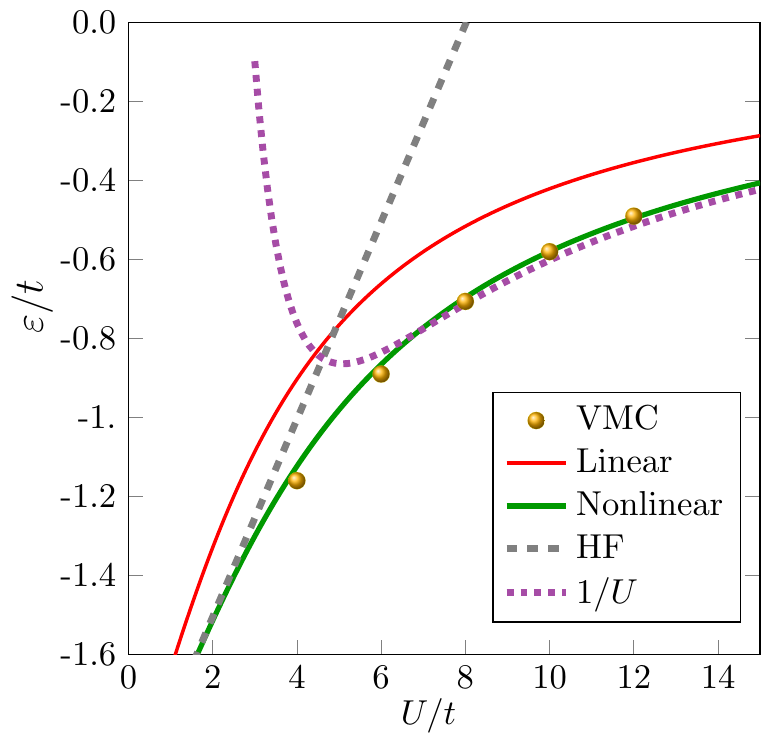}}
\caption{\label{fig3}Comparison of several approximations to the ground-state per-site energy for the three dimensional Hubbard model at half-filling. Colors and symbols as in Fig.~\ref{fig2}. The golden balls denoting VMC calculations come from Ref.~\cite{vmc-shibaII}.}
\end{figure}

\begin{figure}[h]
\centerline{\includegraphics[width=0.95\columnwidth]{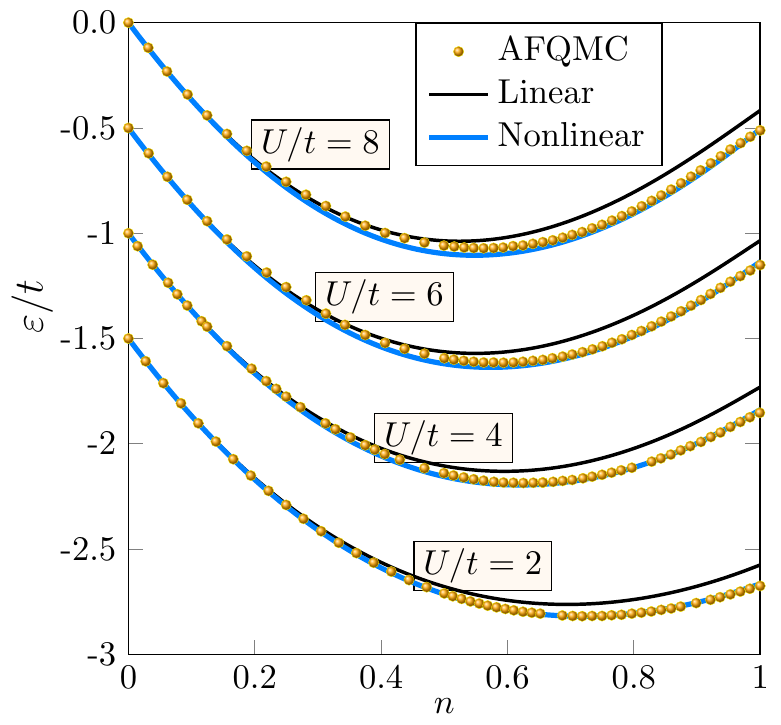}}
\caption{\label{fig4}Comparison between the linear and nonlinear approximations to the per-site ground-state energy for the two-dimensional Hubbard model and AFQMC results from Ref.~\cite{afqmc2008} for the indicated Coulomb repulsions $U$. To avoid crowding, all data for $U/t =2, 4$, and $6$ were shifted vertically by $\Delta\epsilon/t = -1.5, -1$, and $-0.5$, respectively.} 
\end{figure}

Although the linear dimensional scaling in Eq.~(\ref{linear2n}) provides a simple approximation to $\epsilon(d,U,n)$ for $d=2$ and $d=3$, it is clear from Figs.~\ref{fig2}, \ref{fig3} and \ref{fig4} that there is much room for improving accuracy. As in the previous section, we start by considering the half-filled case and then develop an approximation for arbitrary filling.  

\subsection{Half-filled case}
\label{sec:half-filled-case}
Significant improvement results from substituting a more general, polynomial expression for the linear dependence in Eq.~\eqref{linear01}, that is, from writing the expression
\begin{align}
  \epsilon(d,U) \approx
  \epsilon_{\mbox{\tiny NL}}(d,U) = \tau(d,0)\sum_{s=0}^N~\alpha_s x^s. \label{poli}
\end{align}

Since the linear approximation $\epsilon(d,U)\approx \tau(1,U)x(U)$ becomes exact at both the non-interacting ($U=0$, $x=1$) and strongly interacting ($U\to\infty$, $x=0$) limits, a Taylor expansion around either extreme would be unwieldy. It is preferable to treat the $U=0$ and the $U\to\infty$ limits on equal footing. To this end, we will take advantage of exact expressions for $\partial\epsilon/\partial U$ at the $U=0$ and $U\to\infty$ to determine the coefficients $\alpha_2$ and $\alpha_3$ and derive a cubic ($N=3$) form for $\epsilon_{\mbox{\tiny NL}}$. 

As $U\to0$, the Hartree-Fock approximation
\begin{align}
  \label{eq:2}
  \epsilon(d,U\to0,n) = \tau(d,0,n) + \frac{U}{4}n^2
\end{align}
becomes asymptotically exact.

For $d=1$, Eq.~\eqref{eq:2} can be combined with the definition~\eqref{eq:8} to yield the expression
\begin{align}
  \label{eq:4}
  \dfrac{U}{4}n^2 = (x-1)\tau(1,0).
\end{align}

Substitution of the right-hand side of Eq.~\eqref{eq:4} for $U/4$ on the right-hand side of Eq.~\eqref{eq:2} then yields the relation
\begin{align}
  \label{eq:5}
  \epsilon(d,U\to0) = \tau(d,0) + (x-1)\tau(1,0),
\end{align}
which determines the slope at $x=1$ ($U=0$):
\begin{align}
\left.\frac{\partial \epsilon(d,U)}{\partial x}\right|_{x=1} = \tau(1,0), \label{dU_0}
\end{align}
a result that is valid for all $n$.

To determine the large $U$ slope, we recall that, in this limit, the half-filled Hubbard model maps onto the $S=1/2$ antiferromagnetic Heisenberg model with effective interaction $J=4/U$. An accurate asymptotic expansion then describes the ground-state energy~\cite{takahashi77}:
\begin{align}
\epsilon(d,U\gg t) = C_1(d)\frac{t}{U} + C_3(d)\frac{t^3}{U^3} + O\Big(\frac{t^5}{U^5}\Big), \label{largeU}
\end{align}
where $C_1(1) = -4\ln(2)$, $C_3(1) = 9\zeta(3)$, where $\zeta$ is Riemann's zeta function,  $C_1(2) = -4.631788$, $C_3(2) = 34.599816$, $C_1(3) = -6.582948$, and $C_3(3) = 56.612880$.

\begin{table}[!ht]
  \centering
  \caption[Coefficients]{Expansion coefficients on the right-hand side of Eq.~\eqref{largeU} \cite{takahashi77}. For convenient reference, the last column shows the $U=0$ per-site kinetic energies, in units of $t$.}
  \label{tab:0}
  \begin{tabular}{{rlll}}
    \br
    $d$ & \multicolumn{1}{c}{$C_1(d)$} & \multicolumn{1}{c}{$C_3(d)$} & \multicolumn{1}{c}{$\tau(d,0)/t$} \\[7pt]\hline
    1 & $-4\ln(2)$ & $9\zeta(3)$ &   $-4/\pi^{\rm a}$\\
    2 & $-4.631788$ & $34.59816$ & $-1.6211^{\rm b}$\\
    3 & $-6.582948$ & $56.61288$& $-2.0048^{\rm c}$\\
    \br
  \end{tabular}
  \begin{description}
  \item[a] From Ref.~\cite{vmc-shibaII}.
  \item[b] From Ref.~\cite{1989HeV9044}.
  \item[c] From Ref.~\cite{1990Has9168}.
  \end{description}
\end{table}
Equation~(\ref{largeU}) can be rewritten in the form
\begin{align}
  \label{eq:6}
  \epsilon(d,U\gg t,n=1) = \tau(1,0,n=1)\frac{C_1(d)}{C_1(1)}x + O(x^3),
\end{align}
which implies that
\begin{align}
\left.\frac{\partial \epsilon(d,U)}{\partial x}\right|_{x=0} = \tau(1,0)\frac{C_1(d)}{C_1(1)}. \label{dU_inf}
\end{align}

From the exact values of $\epsilon(d,U)$ and its slope at $x=0$ and $x=1$, we determine the coefficients $\alpha_s$ in Eq.(\ref{poli}):
\begin{align}
\alpha_0 &= 0,\label{alp0}\\
\alpha_1 &= \frac{\tau(1,0)}{\tau(d,0)}\frac{C_1(d)}{C_1(1)},\label{alp1}\\
\alpha_2 &= 3 - \frac{\tau(1,0)}{\tau(d,0)} - 2\alpha_1,\label{alp2} \\
\alpha_3 &= -2 +\frac{\tau(1,0)}{\tau(d,0)} + \alpha_1.\label{alp3}
\end{align}

Figures~\ref{fig2}~and \ref{fig3} show that the cubic expression yields excellent agreement with QMC results for the half-filled Hubbard model in 2 and 3 dimensions,  for all $U$. The mapping between the $[0,\infty)$ range of $U$ and the finite range $(0,1]$ of the scaling variable $x(U)$ has allowed us to exploit exact results at both extremes, to obtain a low-order polynomial yielding accurate results over the entire $U$ axis.

In the two-dimensional case, the deviations relative to the AFQMC data shrink from 8.6--22.4\% for the linear to 1--3.6\% for the cubic approximation. In three dimensions, the drop is from 22-27\% to 0.1--3.2\%. Moreover, the procedure we have described can be systematically improved: with higher order terms on the right-hand side of Eq.~\eqref{largeU} as input, one can extend the algebra leading to Eqs.~\eqref{alp0}-\eqref{alp3} to obtain higher order coefficients $\alpha_s$ ($s>3$) and increment the degree $N$ of the polynomial describing $\epsilon(d,U)$.
 
\subsection{Arbitrary filling}
\label{sec:arbitrary-filling-1}
The significant improvement afforded by nonlinear dimensional scaling at half-filling prompts us to generalize the approach in Sec.~\ref{sec:half-filled-case} to arbitrary filling. The following discussion leads to an approximate expression for the ground-state energy at any dimensionality. Unfortunately, due to the scarcity of accurate numerical data away from half-filling for the $3$-dimensional model, we can only check the accuracy of our expression for $d=2$.

To extend the analysis in Sec.~\ref{secnlds} to $n<1$ we simply substitute $\tau(1,0,n)$ for $\tau(1,0)$, and $\tau(d,0,n)$ for $\tau(d,0)$ in Eqs.~\eqref{eq:5}~and \eqref{eq:6} and in Eqs.~\eqref{alp0}-\eqref{alp3}. Again to third order, we then have that
\begin{align}
\epsilon_{\mbox{\tiny NL}}(d,U,n)= \tau(d,0,n)\sum_{s=0}^3~\alpha_s(n) \big(x(U,n)\big)^s, \label{poli-nI}
\end{align}
where the $\alpha_s(n)$ ($s=0,\ldots,3$) are the coefficients given by Eqs.~(\ref{alp0})--(\ref{alp3}) after the aforementioned substitutions.

As explained in Sec.~\ref{sec:arbitrary-filling}, for $n\le1$ the scaling variable $x$ is constrained to the interval $\cos\left(\frac{\pi n}{2}\right)\leq x(U,n) \leq 1$. By construction, both the per-site energy and its derivative $\partial\epsilon(d,U,n)/\partial x$ become exact at $x=1$, for all $n$. In contrast, at the minimum $x=\cos\left(\frac{\pi n}{2}\right)$ ($U\to\infty$), neither the per site energy, nor the derivative resulting from the same approximation are guaranteed to yield exact results. There are two exceptions, however: (i) at half-filling, we recover the results described in Sec.~\ref{sec:half-filled-case}, and (ii) as $n\to0$, the domain of $x$ reduces to a single point, $x=1$, for which the approximation $\epsilon(1,d,U)\approx \epsilon_{\mbox{\tiny NL}}(d,U,n)$ becomes exact.

Figure~\ref{fig4} ratifies these considerations. The agreement between the blue solid line representing the nonlinear approximation and the golden balls representing the AFQMC data \cite{afqmc2008} is excellent in the vicinity of the extremes, $n=0$ and $n=1$, while the small deviations around $n=0.5$ become more clearly visible as $U/t$ grows.

\begin{table}
  \caption{\label{table1} Per site ground-state energy for the two-dimensional model resulting from the analyses in Secs.~\ref{sec:arbitrary-filling}~and \ref{sec:arbitrary-filling-1}, for several Coulomb repulsions $U$ and densities $n$. The rightmost column shows the benchmarks set by the available Quantum Monte Carlo. Over the subset of values of $U$ and $n$ for which QMC data are available, the mean average error of the linear scaling approximation with respect to the QMC values is 10.5\%. The nonlinear generalization, truncated after the cubic term, reduces this to 3.14\%.}
  \begin{tabular}{ccccc}\br
  $U$  & $n$& Linear scaling & Nonlinear scaling & QMC \\\hline
  3  &	0.25  &	-0.779  &	-0.793  & \\
  3  &	0.50  &	-1.137  &	-1.181  & \\
  3  &	0.75  &	-1.150  &	-1.225  & \\
  3  &	0.90625  &	-1.011  &	-1.105  & \\
  3  & 1.00  & -0.879  & -0.983  & \\\hline
  5  &	0.25  &	-0.771  &	-0.788  & \\
  5  &	0.50  &	-1.082  &	-1.136  & \\
  5  &	0.75  &	-1.003  &	-1.093  & \\
  5  &	0.90625  &	-0.790  &	-0.896  & \\  
  5  & 1.00  & -0.619  & -0.725  & \\\hline
  6  & 0.25  & -0.769  & -0.786  & -0.756$^{\rm a}$ \\	
  6  & 0.50  & -1.063  & -1.120  & -1.093$^{\rm a}$ \\
  6  & 0.75  & -0.952  & -1.046  & -1.037$^{\rm a}$ \\
  6  & 0.90625  & -0.717 & -0.823 & -0.817$^{\rm a}$ \\	
  6  & 1.00  & -0.535  & -0.635  & -0.651$^{\rm a}$ \\\hline
  10  &	0.25  &	-0.764  &	-0.782  & \\
  10  &	0.50  &	-1.018  &	-1.081  & \\
  10  &	0.75  &	-0.829  &	-0.927  & \\
  10 & 0.78  & -0.783  & -0.883 & -0.843$^{\rm b}$ \\
  10 & 0.89  & -0.578  & -0.679 & -0.644$^{\rm b}$ \\  
  10 & 0.94  & -0.472  & -0.566 & -0.554$^{\rm b}$ \\\hline
  12  &	0.25  &	-0.762  &	-0.781  & \\
  12  &	0.50  &	-1.005  &	-1.069  & \\
  12  &	0.75  &	-0.794  &	-0.892  & \\
  12 & 0.80  & -0.709  & -0.810  & -0.772$^{\rm c}$ \\
  12 & 0.875 & -0.562  & -0.662  & -0.628$^{\rm c}$ \\
  12  &	0.90625  &	-0.496  &	-0.591  & \\  
    12 & 1.00  & -0.287  & -0.356  & -0.369$^{\rm c}$ \\
    \br
  \end{tabular}
  \begin{description}
  \item[] $^{\rm a}$ From AFQMC calculation in Ref.~\cite{afqmc2008}.
  \item[] $^{\rm b}$ From Variational Monte Carlo calculation in Ref.~\cite{vmc2017}.
  \item[] $^{\rm c}$ From AFQMC calculation in Ref.~\cite{prx15}.
  \end{description}
\end{table}

The same reasoning shows that, at low densities, the nonlinear and linear approximations become equivalent. At half-filling, by contrast, the corrections to the linear approximation defined by the terms proportional to $\alpha_2$ and $\alpha_3$ on the right-hand side of Eq.~\eqref{poli-nI} are sizeable, even for $U$ as large as $8t$. 

The AFQMC data taken as benchmarks, Fig.~\ref{fig4} shows that linear approximation overestimates the per-site ground-state energy, while the nonlinear approximation underestimates it. The same trend is apparent in Table~\ref{table1}, which displays more data for the two-dimensional model. A number of results without matching  benchmarks are shown, to encourage additional QMC computations. With the same motivation, we present in Table~\ref{table2} results for the three-dimensional model, for which we have been unable to find QMC results representative of the thermodynamic limit away from half-filling.

\begin{table}
  \caption{\label{table2} Per site ground-state energy for the three-dimensional model resulting from the analyses in Secs.~\ref{sec:arbitrary-filling}~and \ref{sec:arbitrary-filling-1}, for several Coulomb repulsions $U$ and densities $n$. Benchmarks set by the available Quantum Monte-Carlo data are tabulated.
  Over the subset of values of $U$ and $n$ for which QMC data are available, the mean average error of the linear scaling approximation with respect to the QMC values is 26\%. The nonlinear generalization, truncated after the cubic term, reduces this to 1.65\%.}
\begin{tabular}{ccccc}
\br

$U$  & $n$&    \begin{tabular}{c}linear scaling\\prediction\end{tabular} & \begin{tabular}{c}nonlinear scaling\\prediction\end{tabular} & QMC \\\hline
4.0  & 0.25  &	-0.954  &	-0.978  & \\
4.0  & 0.50  &	-1.320  &	-1.407  & \\
4.0  & 0.75  &	-1.299  &	-1.466  & \\
4.0  & 0.90625  &	-1.092  &	-1.303  & \\
4.0  & 1.00  &	-0.903  &	-1.124  & -1.16$^{\rm a}$\\\hline
6.0  & 0.25  &	-0.947  &	-0.974  & \\
6.0  & 0.50  &	-1.269  &	-1.369  & \\
6.0  & 0.75  &	-1.159  &	-1.346  & \\
6.0  & 0.90625  &	-0.883  &	-1.101  & \\
6.0  & 1.00  &	-0.661  &	-0.866  & -0.89$^{\rm a}$\\\hline
8.0  & 0.25  &	-0.943  &	-0.972  & \\
8.0  & 0.50  &	-1.237  &	-1.345  & \\
8.0  & 0.75  &	-1.069  &	-1.265  & \\
8.0  & 0.90625  &	-0.755  &	-0.966  & \\
8.0  & 1.00  &	-0.516  &	-0.696  & -0.706$^{\rm a}$\\\hline
10.0  & 0.25  &	-0.940  &	-0.970  & \\
10.0  & 0.50  &	-1.215  &	-1.328  & \\
10.0  & 0.75  &	-1.009  &	-1.209  & \\
10.0  & 0.90625  &	-0.670  &	-0.874  & \\
10.0  & 1.00  &	-0.421  &	-0.580  & -0.58$^{\rm a}$\\\hline
12.0  & 0.25  &	-0.9384  &	-0.969  & \\
12.0  & 0.50  &	-1.1996  &	-1.315  & \\
12.0  & 0.75  &	-0.9657  &	-1.167  & \\
12.0  & 0.90625  &	-0.6112  &	-0.807  & \\
12.0  & 1.00  &	-0.3544  &	-0.495  & -0.49$^{\rm a}$\\
\br
\end{tabular}
\begin{description}
\item[] $^{\rm a}$ {From the Variational Monte Carlo calculation in Ref.~\cite{vmc-shibaII}}
\end{description}
\end{table}

The deviations separating the blue lines from the golden balls in Fig.~\ref{fig4} leave room for further improvement of the nonlinear scaling procedure. Unfortunately, progress along that line of work calls for exact results for the $U\to\infty$ limits of the ground-state energy and its derivative with respect to $U$ for $d>1$ and $n\ne 1$, which are unavailable. We have tested a nonlinear scaling approximation for $d=2$ and $n<1$ built upon Eq.~(\ref{euinfty}), which offers an (uncontrollable) approximation for the per-site ground-state energy. The resulting expression for $\epsilon_{\mbox{\tiny NL}}(2,U, n)$ is not shown because it is inferior to the one represented by the blue lines in Fig.~\ref{fig4}.

\section{Conclusion}
\label{sec:conclusion}

We have presented a rationale justifying, for the first time, an approximation of capital importance in the dimensional scaling approaches introduced in Refs.~\cite{LK2006}~and \cite{akandeBJP}: the substitution of the noninteracting kinetic energy for the (unknown) interacting kinetic energy on the right-hand side of Eq.~\eqref{linear0}, which defines the scaling variable.

The same line of reasoning led to a nonlinear approximation for the per-site ground-state energy. Exact results for the per-site ground-state energy for the two- and three-dimensional Hubbard models at half-filling gave access to the energy and its derivative with respect to the Coulomb repulsion $U$ in the noninteracting ($U=0$) and strongly interacting ($U\to\infty$) limits. The resulting cubic approximations for the per-site ground-state energies, devoid of adjustable parameters, yielded substantial improvement over the agreement with accurate Quantum Monte Carlo data for the two- and three-dimensional models at half-filling. 

Away from half-filling, a region of the parametric space of the two- and three-dimensional models for which exact results are unavailable in the $U\to\infty$ limit, the agreement between our results and the Quantum Monte-Carlo data for the two-dimensional model is still good. Tables~\ref{table1}~and \ref{table2} displays results generated by the linear and nonlinear scaling approaches, for a number of ($n$,$U$) pairs, with a view to motivating additional Monte-Carlo computations for the two- and three-dimensional models. In the absence of such QMC calculations, our results constitute predictions for the ground-state energy at the reported $(n,U)$ pairs. The simplicity of the cubic scaling expression permits to easily generate further data for other $(n,U)$ pairs in two and three dimensions.

In addition to providing predictions that can be checked by QMC, the availability of a relatively simple explicit expression for the ground-state energy as a function of $n$ and $U$ opens up the possibility of an occupation-number functional approach~\cite{physrepKCVC} to the inhomogeneous Hubbard model in $d>1$, within the local-density approximation,  which could treat finite lattices of large sizes, subject to arbitrary external potentials. We plan to explore this application of the dimensional scaling approach in future work.

On the conceptual side, it remains a challenge to verify if the nonlinear expression for the Hubbard model can be further improved, by incorporating additional exact properties into higher-order terms, or exploiting the variational principle. More generally, it would be rewarding to search for further contraints, similar to the virial theorem for the Fermi liquid and the Hellmann-Feynman theorem for the Hubbard model, that enable interdimensional scaling laws for other systems and physical quantities.

\section*{Acknowledgments}
\label{sec:acknowledgments}
This work was supported by the CNPq (grant number 312658/2013-3).

\appendix

\section{Inequivalence between the linear approximations}
\label{sec:ineq-betw-line}
To show that Eqs.~\eqref{linear0}~and \eqref{linear01} are inequivalent, we examine both equalities for large $U$. In that domain, the ground state energy is given by Eq.~\eqref{largeU}. The corresponding kinetic energy can be obtained from Eq.~\eqref{eq:9}:
\begin{align}
  \label{eq:13}
  \tau(d,U\gg t) =2C_1(d) \frac{t}{U} + 4C_3(d)\frac{t^3}{U^3}
  +O\Big(\frac{t^5}{U^5}\Big),
\end{align}
and therefore
\begin{align}
  \label{eq:14}
  \dfrac{\epsilon(d,U\gg t)}{\tau(d,U)} =\frac12+ O\Big(\frac{t^2}{U^2}\Big),
\end{align}
while
\begin{align}
  \label{eq:15}
   \dfrac{\epsilon(d,U\gg t)}{\tau(d,0)} =\frac{C_1(d)}{\tau(d,0)} \frac{t}{U} + \frac{C_3(d)}{\tau(d,0)}\frac{t^3}{U^3}
  +O\Big(\frac{t^5}{U^5}\Big).
\end{align}

Substitution of the coefficients and kinetic energies in Table~\ref{tab:0} on the right-hand side of Eq.~\eqref{eq:15} now yields the following expressions:
\begin{align}
  \label{eq:16}
\dfrac{\epsilon(1,U\gg t)}{\tau(1,0)} =\pi\ln(2)\frac{t}{U} +O\Big(\frac{t^3}{U^3}\Big),
\end{align}
\begin{align}
  \label{eq:17}
  \dfrac{\epsilon(2,U\gg t)}{\tau(d,0)} =2.8572 \frac{t}{U} +O\Big(\frac{t^3}{U^3}\Big),
\end{align}
and
\begin{align}
  \label{eq:18}
  \dfrac{\epsilon(3,U\gg t)}{\tau(d,0)} =3.2836 \frac{t}{U} +O\Big(\frac{t^3}{U^3}\Big).
\end{align}

We can now compare Eq.~\eqref{eq:14}, which states that, to $O(t/U)$, the ratio $\epsilon(d,U\gg t)/\tau(d,U)$ is a constant, independent of $d$, with Eqs.~\eqref{eq:16}-\eqref{eq:18}, which show that the linear coefficient in the $t/U$ expansion varies with $d$. This distinction is sufficient to show that Eqs.~\eqref{linear0}~and \eqref{linear01} cannot be equivalent.

\section*{References}
\label{sec:references}

\bibliography{Refs}
\bibliographystyle{unsrt}
\end{document}